\documentclass[aps,prb,reprint,amsfonts,amsmath,amssymb,longbibliography,showkeys]{revtex4-2}
\usepackage{hyperref}
\usepackage{graphicx}
\usepackage{bm}
\usepackage{newtxtext}
\usepackage[varg]{newtxmath}
\usepackage[normalem]{ulem}
\usepackage{braket}
\hypersetup{colorlinks=true,linkcolor=blue,citecolor=blue,urlcolor=blue}
\usepackage{mathtools}
\usepackage{mleftright}
\mleftright

\DeclareMathOperator*{\Tr}{Tr}

\allowdisplaybreaks

\begin{document}
\title{Temporal modulation of second harmonic generation in ferroelectrics by a pulsed electric field}
\author{\mbox{Atsushi Ono}}
\affiliation{\mbox{Department of Physics, Graduate School of Science, Tohoku University, Sendai 980-8578, Japan}}
\date{September 25, 2025}

\begin{abstract}
We revisit the relationship between electric polarization modulation $\Delta P$ in ferroelectrics induced by a low-frequency pulsed electric field and the corresponding second harmonic intensity modulation $\Delta I_{\mathrm{SH}}$.
Using nonlinear response theory, we derive their time-domain expressions linear in the pulse amplitude, revealing that not only the electric field but also its time derivative contributes to $\Delta I_{\mathrm{SH}}$.
Furthermore, even when the time-derivative component is negligible, $\Delta I_{\mathrm{SH}}$ can be in antiphase with the pulsed field and thus with $\Delta P$.
These theoretical predictions are further supported by real-time simulations of a model for electronic ferroelectrics.
Our results demonstrate that the commonly assumed relation $\Delta P \propto \Delta I_{\mathrm{SH}}$ can break down under certain conditions, reflecting the complex-valued and frequency-dependent nature of nonlinear dynamical susceptibilities.
\end{abstract}

\makeatletter
\renewcommand{\@keys@name}{DOI: }
\makeatother
\keywords{\href{https://doi.org/10.1103/xsrf-t1hj}{10.1103/xsrf-t1hj}}

\maketitle

\section{Introduction} \label{sec:intro}
Second harmonic generation (SHG) emerged almost simultaneously with the invention of the laser, when Franken \textit{et al.}\ observed light at twice the frequency of a ruby laser beam transmitted through quartz \cite{Franken1961}.
A year later, the perturbation formalism and phase-matching rules were rigorously formulated \cite{Armstrong1962}, establishing SHG as a second-order optical process that is forbidden in centrosymmetric crystals and thus sensitive to inversion symmetry breaking \cite{Franken1963, Shen1989, Fiebig2005, Ma2020, Kirilyuk2010}.
Since then, SHG has evolved into a nondestructive probe of ferroelectricity and ferroelectric domains \cite{Kurtz1968, Uesu1995, Trassin2015, Zhang2018p, Spychala2020, Fujiwara2021, Abdelwahab2022, Hegarty2022}; a symmetry probe for multiferroics \cite{Fiebig2016, Lottermoser2004, Sharan2004, Jin2020k, Toyoda2021a, Xu2023, Prasad2024} and magnets \cite{Nemec2018, Hohlfeld1997, Fiebig2001, Pavlov2005, Tzschaschel2019, Okumura2021, Xiao2023a, Toyoda2023, Shoriki2024, Wu2024e, Wu2025a}; a valley-selective spectroscopic tool in two-dimensional semiconductors \cite{Kumar2013, Li2013, Seyler2015, Herrmann2025}; and a probe of quantum geometry \cite{Orenstein2021, Hu2025, Morimoto2016, Wu2017, Li2018, Patankar2018, Bhalla2022, Watanabe2022, Tanaka2024, Tzschaschel2024, He2021, Sodemann2015, Okyay2022, Denisov2025}.

More recently, the advent of intense femtosecond and terahertz pulses has broadened interest from static to dynamic symmetry control.
Combined with optical pulse excitation, time-resolved SHG spectroscopy enables a direct readout of transient ferroelectric order in quantum paraelectrics \cite{Li2019br, Nova2019, Shin2022, Li2023, Bilyk2023, Yang2025} and ferroelectrics \cite{Miyamoto2013, Yamakawa2016b, Iwano2017, Morimoto2017b, Miyamoto2018b, Yamakawa2021, Kida2022, Umanodan2019, Sugisawa2023, Mankowsky2017, Grishunin2017, Grishunin2019, Chen2015, Yu2020p, Yu2024, Takubo2024, Itoh2025}, as well as other ferroic orders in magnets and multiferroics \cite{Sheu2016, Matsubara2019a, BustamanteLopez2025}.
This technique also probes inversion symmetry breaking in diverse nonequilibrium systems \cite{Terhune1962, Sie2019, Vaswani2019, Ovchinnikov2019, Tokman2019, Kawakami2020, Takasan2021, Gao2021b, Nakamura2020, Nakamura2024a, Wang2024}.
In the quest for dynamic control of symmetry and order in condensed matter, SHG now plays an increasingly pivotal role.

To analyze SHG signals, the following phenomenological expression has been widely used not only under equilibrium conditions but also in nonequilibrium photoexcited states:
\begin{align}
P_{i}(2\omega) = \sum_{jk} \tilde{\chi}_{ijk}^{(2)}(2\omega; \omega, \omega) E_{j}(\omega) E_{k}(\omega).
\label{eq:Pw_phenom}
\end{align}
Here, $E_{i}(\omega)$ and $P_{i}(2\omega)$ denote the $i$-component of the probe electric field at frequency $\omega$ and the induced polarization at frequency $2\omega$, respectively, and $\tilde{\chi}_{ijk}^{(2)}$ is the second-order susceptibility tensor.
The intensity of SHG is proportional to the square of $P_{i}(2\omega)$.
The term ``phenomenological'' is used here because $\tilde{\chi}_{ijk}^{(2)}$ coincides with the microscopic response function derived from time-dependent perturbation theory only when the probe electric field is strictly monochromatic (i.e., a continuous wave).

Even when the probe is approximately monochromatic, additional approximations are often made in discussing the polarization modulation $\Delta P$ induced by a pump pulse $E_{\mathrm{pump}}$ (indices omitted for simplicity).
It is commonly assumed that the second-order susceptibility changes as $\tilde{\chi}^{(2)} \to \tilde{\chi}^{(2)} + \Delta \tilde{\chi}^{(2)}$, with $\Delta \tilde{\chi}^{(2)} \propto E_{\mathrm{pump}}$, and hence that the resulting change in the second harmonic intensity is
\begin{align}
\Delta I_{\mathrm{SH}} \propto \bigl[ \tilde{\chi}^{(2)} + \Delta \tilde{\chi}^{(2)} \bigr]^2 - \bigl[ \tilde{\chi}^{(2)} \bigr]^2 \approx 2 \tilde{\chi}^{(2)} \Delta \tilde{\chi}^{(2)} \propto E_{\mathrm{pump}}.
\end{align}
Since $\Delta P \propto E_{\mathrm{pump}}$, one then infers that $\Delta P \propto \Delta I_{\mathrm{SH}}$.
However, $\Delta \tilde{\chi}^{(2)}$ should, in principle, be evaluated as a convolution of the third-order susceptibility with the full spectra of the probe and pump fields [see Eq.~\eqref{eq:Pw}], so it is not obvious that $\Delta \tilde{\chi}^{(2)}$ remains strictly in phase with $E_{\mathrm{pump}}$.
This subtlety appears to have received little attention in the literature, except for a pioneering study by Timm and Bennemann \cite{Timm2004}.

In this paper, we revisit SHG that is temporally modulated by a pulsed electric field whose frequency is much lower than the probe frequency.
In Sec.~\ref{sec:theory}, we derive a general expression for $\Delta I_{\mathrm{SH}}$ based on microscopic nonlinear response theory.
In Sec.~\ref{sec:MF}, we validate this expression using real-time simulations of a model ferroelectric system and demonstrate cases in which the relation $\Delta P \propto \Delta I_{\mathrm{SH}}$ fails.
Sections~\ref{sec:discussion} and \ref{sec:summary} provide the discussion and summary, respectively.
In Appendix, we present the spectral representation of the nonlinear susceptibility used for a quantitative comparison between theory and simulation.

\section{General considerations} \label{sec:theory}
We consider the Hamiltonian
\begin{align}
\mathcal{H}(t) = \mathcal{H}_0 + \mathcal{V}(t),
\end{align}
where $t$ is time, $\mathcal{H}_0$ is the unperturbed Hamiltonian, and $\mathcal{V}(t)$ is the time-dependent perturbation:
\begin{align}
\mathcal{V}(t) = - E(t) P,
\label{eq:perturbation}
\end{align}
with $E(t)$ the external electric field and $P$ the electric polarization operator.
The expectation value of the electric polarization is given by
\begin{align}
\langle P \rangle (t) = \Tr[\rho(t) P],
\end{align}
where $\rho(t)$ is the density matrix of the system.

According to time-dependent perturbation theory, the $n$th-order response of the polarization to the electric field is expressed as
\begin{align}
\langle P \rangle^{(n)}(t)
&= \int_{-\infty}^{\infty} \mathrm{d}t_1 \cdots \mathrm{d}t_n\, \chi^{(n)}(t;t_1,\dots,t_n) E(t_1)\cdots E(t_n).
\label{eq:Pn_t}
\end{align}
Here, the $n$th-order susceptibility $\chi^{(n)}$ is defined by
\begin{align}
&\chi^{(n)}(t;t_1,t_2,\dots,t_n) \notag \\
&= \mathrm{i}^n \varTheta(t-t_1)\varTheta(t_1-t_2)\cdots\varTheta(t_{n-1}-t_n) \notag \\
&\quad\times \left\langle [[\cdots [[\hat{P}(t),\hat{P}(t_1)],\hat{P}(t_2)],\cdots ], \hat{P}(t_n)] \right\rangle_0,
\label{eq:chi_t}
\end{align}
where $\langle \bullet \rangle_0 = \Tr[\rho(-\infty) \bullet ]$ is the expectation value in the initial state, $\varTheta$ is the unit step function, and $\hat{P}$ is the interaction-picture operator defined by
\begin{gather}
\hat{P}(t) = \mathcal{U}_0(t, -\infty)^\dagger P \mathcal{U}_0(t, -\infty), \\
\mathcal{U}_0(t, -\infty) = \mathcal{T} \exp\left( -\mathrm{i} \int_{-\infty}^{t} \mathcal{H}_0\, \mathrm{d}t' \right),
\end{gather}
with $\mathcal{T}$ the time-ordering operator.
Throughout this paper, we set the Dirac constant $\hbar = 1$.
Since Eq.~\eqref{eq:chi_t} is invariant under time translation,
\begin{align}
\chi^{(n)}(t;t_1,\dots,t_n) = \chi^{(n)}(t+T;t_1+T,\dots,t_n+T)
\end{align}
for any $T$, we introduce an $n$-variable susceptibility:
\begin{align}
\chi^{(n)}(\bar{t}_1,\dots,\bar{t}_n)
= \chi^{(n)}(0; -\bar{t}_1, \dots, -\bar{t}_n),
\end{align}
where $\bar{t}_i = t-t_i$ for $i=1$ to $n$.
The $n$th-order susceptibility in the frequency domain is given by
\begin{align}
&\chi^{(n)}(\omega_1,\dots,\omega_n) \notag \\
&= \int_{-\infty}^{\infty} \mathrm{d}\bar{t}_1 \cdots \mathrm{d}\bar{t}_n\, \mathrm{e}^{\mathrm{i}\omega_1^+\bar{t}_1} \cdots \mathrm{e}^{\mathrm{i}\omega_n^+\bar{t}_n} \chi^{(n)}(\bar{t}_1,\dots,\bar{t}_n), \label{eq:chi_w}
\end{align}
with $\omega_i^+ = \omega_i + \mathrm{i}\eta$, where $\eta > 0$ is a broadening factor.
Since $\hat{P}$ is Hermitian, $\chi^{(n)}$ in Eq.~\eqref{eq:chi_w} satisfies the relation
\begin{align}
\chi^{(n)}(\omega_1,\dots,\omega_n) = \chi^{(n)}(-\omega_1,\dots,-\omega_n)^*.
\label{eq:hermitian}
\end{align}
The Fourier transformation of $\langle P \rangle^{(n)}(t)$ yields
\begin{align}
\langle P \rangle^{(n)}(\omega) &= \int_{-\infty}^{\infty} \mathrm{d}t\, \mathrm{e}^{\mathrm{i} \omega t} \langle P \rangle^{(n)}(t) \\
&= \int_{-\infty}^{\infty} \frac{\mathrm{d}\omega_1 \cdots \mathrm{d}\omega_n}{(2\pi)^{n-1}} \delta(\omega_1 + \cdots + \omega_n - \omega) \notag \\
&\quad \times \chi^{(n)}(\omega_1, \dots, \omega_n) E(\omega_1) \cdots E(\omega_n),
\label{eq:Pw}
\end{align}
where $\delta$ represents the delta function.
Unlike Eq.~\eqref{eq:Pw_phenom}, Eq.~\eqref{eq:Pw} is valid for arbitrary $E(\omega)$.

In what follows, we discuss the temporal modulations in the polarization and the second harmonic (SH) intensity induced by the following external electric field:
\begin{align}
E(t) = E_{\mathrm{IR}}\cos(\omega_{\mathrm{IR}}t) + E_{\mathrm{THz}}(t), \label{eq:Et}
\end{align}
where the first and second terms represent an infrared (IR) probe and a terahertz (THz) pump, respectively.
In the frequency domain, the electric field is expressed as
\begin{align}
E(\omega)
&= E_{\mathrm{IR}} \pi [\delta(\omega-\omega_{\mathrm{IR}}) + \delta(\omega+\omega_{\mathrm{IR}})] + E_{\mathrm{THz}}(\omega). \label{eq:Ew}
\end{align}

\subsection{Electric polarization}
Setting $E_{\mathrm{IR}} = 0$ and substituting Eq.~\eqref{eq:Ew} into Eq.~\eqref{eq:Pw}, we obtain the well-known expression for the polarization change linear in $E_{\mathrm{THz}}$ as follows:
\begin{align}
\langle P \rangle^{(1)}(\omega)
= \chi^{(1)}(\omega) E_{\mathrm{THz}}(\omega).
\end{align}
When $E_{\mathrm{THz}}(\omega)$ is sharply peaked around $\omega = 0$, $\chi^{(1)}(\omega)$ can be approximated by its value at $\omega = 0$, yielding
\begin{align}
\langle P \rangle^{(1)}(\omega)
\approx \chi^{(1)}(\omega = 0) E_{\mathrm{THz}}(\omega).
\end{align}
Since $\chi^{(1)}(\omega = 0)$ is real-valued, as follows from Eq.~\eqref{eq:hermitian}, there is no phase delay between the polarization and the pulsed field.
Accordingly, in the time domain, we arrive at the relation
\begin{align}
\Delta P(t)
\equiv \langle P \rangle^{(1)}(t)
\propto E_{\mathrm{THz}}(t).
\label{eq:Pt_modulation}
\end{align}

\subsection{Second harmonic generation}
We now consider the second-order response of the IR probe under the THz pump pulse described by Eq.~\eqref{eq:Et}.
For the response in the absence of the pump pulse, by substituting Eq.~\eqref{eq:Ew} into Eq.~\eqref{eq:Pw} and extracting the dominant contributions near $\omega = \pm 2 \omega_{\mathrm{IR}}$, we obtain
\begin{align}
P^{(2)}(\omega)
&= \frac{\pi}{4} \bar{\chi}^{(2)}(\omega_{\mathrm{IR}},\omega_{\mathrm{IR}}) E_{\mathrm{IR}}^2 \delta(\omega-2\omega_{\mathrm{IR}}) \notag \\
&\quad + \frac{\pi}{4} \bar{\chi}^{(2)}(-\omega_{\mathrm{IR}},-\omega_{\mathrm{IR}}) E_{\mathrm{IR}}^2 \delta(\omega+2\omega_{\mathrm{IR}}).
\label{eq:Pw2}
\end{align}
Similarly, the pump-induced response linear in $E_{\mathrm{THz}}$ can be expressed in terms of the third-order susceptibility as 
\begin{align}
&P^{(3)}(\omega) \notag \\
&= \frac{1}{8} \bar{\chi}^{(3)}(\omega_{\mathrm{IR}},\omega_{\mathrm{IR}},\omega-2\omega_{\mathrm{IR}}) E_{\mathrm{IR}}^2 E_{\mathrm{THz}}(\omega-2\omega_{\mathrm{IR}}) \notag \\
&\quad + \frac{1}{8} \bar{\chi}^{(3)}(-\omega_{\mathrm{IR}},-\omega_{\mathrm{IR}},\omega+2\omega_{\mathrm{IR}}) E_{\mathrm{IR}}^2 E_{\mathrm{THz}}(\omega+2\omega_{\mathrm{IR}}). \label{eq:Pw3}
\end{align}
In Eqs.~\eqref{eq:Pw2} and \eqref{eq:Pw3}, we define the symmetrized susceptibility as
\begin{align}
\bar{\chi}^{(n)}(\omega_1, \dots, \omega_n)
= \sum_{s \in \mathfrak{S}_n} \chi^{(n)}(\omega_{s(1)}, \dots, \omega_{s(n)}),
\label{eq:chi_symmetrized}
\end{align}
where $\mathfrak{S}_n$ represents the symmetric group of degree $n$.

The far field of the electric dipole radiation, which is proportional to the second derivative of $P(t)$, is defined as follows (up to a multiplicative constant):
\begin{align}
E_{\infty}(t) = -\ddot{P}(t), \quad
E_{\infty}(\omega) = \omega^2 P(\omega). \label{eq:dipole_w}
\end{align}
By substituting Eq.~\eqref{eq:Ew} into Eq.~\eqref{eq:dipole_w}, we obtain
\begin{align}
&E_{\infty}(\omega) \notag \\
&= \frac{\pi}{4} E_{\mathrm{IR}}^2
\Bigl[ (2\omega_{\mathrm{IR}})^2 \bar{\chi}^{(2)}(\omega_{\mathrm{IR}},\omega_{\mathrm{IR}}) \delta(\omega-2\omega_{\mathrm{IR}}) \notag \\
&\quad + (-2\omega_{\mathrm{IR}})^2 \bar{\chi}^{(2)}(-\omega_{\mathrm{IR}},-\omega_{\mathrm{IR}}) \delta(\omega+2\omega_{\mathrm{IR}}) \Bigr] \notag \\
&\quad + \frac{1}{8} E_{\mathrm{IR}}^2
\Bigl[ \omega^2 \bar{\chi}^{(3)}(\omega_{\mathrm{IR}},\omega_{\mathrm{IR}},\omega-2\omega_{\mathrm{IR}}) E_{\mathrm{THz}}(\omega-2\omega_{\mathrm{IR}}) \notag \\
&\quad + \omega^2 \bar{\chi}^{(3)}(-\omega_{\mathrm{IR}},-\omega_{\mathrm{IR}},\omega+2\omega_{\mathrm{IR}}) E_{\mathrm{THz}}(\omega+2\omega_{\mathrm{IR}}) \Bigr].
\label{eq:Ew_inf}
\end{align}
To simplify notation, we introduce the following shorthand:
\begin{align}
\bar{\chi}_2 &= \bar{\chi}^{(2)}(\omega_{\mathrm{IR}},\omega_{\mathrm{IR}}), \label{eq:sh_chi2} \\
\bar{\chi}_3 &= \bar{\chi}^{(3)}(\omega_{\mathrm{IR}},\omega_{\mathrm{IR}},0), \label{eq:sh_chi3} \\
\bar{\chi}_3' &= \left. \frac{\partial \bar{\chi}^{(3)}(\omega_{\mathrm{IR}},\omega_{\mathrm{IR}}, \omega-2\omega_{\mathrm{IR}})}{\partial \omega} \right\vert_{\omega = 2\omega_{\mathrm{IR}}}. \label{eq:sh_chi3d}
\end{align}
The third-order susceptibility is then expanded around $\omega = \pm 2 \omega_{\mathrm{IR}}$ as
\begin{align}
\bar{\chi}^{(3)}(\omega_{\mathrm{IR}},\omega_{\mathrm{IR}},\omega-2\omega_{\mathrm{IR}})
&\approx \bar{\chi}_3 + (\omega - 2\omega_{\mathrm{IR}}) \bar{\chi}_3', \label{eq:chi3_shorthand_p} \\
\bar{\chi}^{(3)}(-\omega_{\mathrm{IR}},-\omega_{\mathrm{IR}},\omega+2\omega_{\mathrm{IR}})
&= \bar{\chi}^{(3)}(\omega_{\mathrm{IR}},\omega_{\mathrm{IR}},-\omega-2\omega_{\mathrm{IR}})^* \notag \\
&\approx \bar{\chi}_3^* - (\omega + 2\omega_{\mathrm{IR}}) \bar{\chi}_3'^*, \label{eq:chi3_shorthand_m}
\end{align}
where we have used Eq.~\eqref{eq:hermitian}.
By performing the inverse Fourier transformation of $E_{\infty}(\omega)$ and neglecting second- and higher-order derivatives of $E_{\mathrm{THz}}(t)$, we obtain the time-domain expression for the far field:
\begin{widetext}
\begin{align}
E_{\infty}(t)
&= \bigl\{ \omega_{\mathrm{IR}}^2 \vert \bar{\chi}_2 \vert \cos(\phi_2) + \omega_{\mathrm{IR}}^2 \vert \bar{\chi}_3 \vert \cos(\phi_3) E_{\mathrm{THz}}(t) - \bigl[ \omega_{\mathrm{IR}} \vert \bar{\chi}_3 \vert \sin(\phi_3) + \omega_{\mathrm{IR}}^2 \vert \bar{\chi}_3' \vert \sin(\phi_3') \bigr] \dot{E}_{\mathrm{THz}}(t) \bigr\} E_{\mathrm{IR}}^2 \cos(2\omega_{\mathrm{IR}} t) \notag \\
&\quad + \bigl\{ \omega_{\mathrm{IR}}^2 \vert \bar{\chi}_2 \vert \sin(\phi_2) + \omega_{\mathrm{IR}}^2 \vert \bar{\chi}_3 \vert \sin(\phi_3) E_{\mathrm{THz}}(t) + \bigl[ \omega_{\mathrm{IR}} \vert \bar{\chi}_3 \vert \cos(\phi_3) + \omega_{\mathrm{IR}}^2 \vert \bar{\chi}_3' \vert \cos(\phi_3') \bigr] \dot{E}_{\mathrm{THz}}(t) \bigr\} E_{\mathrm{IR}}^2 \sin(2\omega_{\mathrm{IR}} t),
\label{eq:farfield}
\end{align}
\end{widetext}
where $\phi_2 = \arg \bar{\chi}_2$, $\phi_3 = \arg \bar{\chi}_3$, and $\phi_3' = \arg \bar{\chi}_3'$.

The appearance of the time derivative of $E_{\mathrm{THz}}(t)$ in Eq.~\eqref{eq:farfield} can be attributed to two distinct origins.
The first arises from the fact that the far field $E_{\infty}(t)$ is proportional to the second derivative of the polarization, as shown in Eq.~\eqref{eq:dipole_w}.
Specifically, the term $\omega^2 E_{\mathrm{THz}}(\omega \mp 2\omega_{\mathrm{IR}})$ in Eq.~\eqref{eq:Ew_inf} leads to time derivatives of $E_{\mathrm{THz}}(t)$ through its inverse Fourier transformation:
\begin{align}
&\int_{-\infty}^{\infty} \frac{\mathrm{d}\omega}{2\pi} \mathrm{e}^{-\mathrm{i}\omega t} \omega^2 E_{\mathrm{THz}}(\omega \mp 2\omega_{\mathrm{IR}}) \notag \\
&= - \frac{\mathrm{d}^2}{\mathrm{d}t^2} \int_{-\infty}^{\infty} \frac{\mathrm{d}\omega}{2\pi} \mathrm{e}^{-\mathrm{i}\omega t} \int_{-\infty}^{\infty} \mathrm{d}t' \mathrm{e}^{\mathrm{i}(\omega \mp 2\omega_{\mathrm{IR}})t'} E_{\mathrm{THz}}(t') \notag \\
&= \mathrm{e}^{\mp 2\mathrm{i}\omega_{\mathrm{IR}}t} \bigl[ 4 \omega_{\mathrm{IR}}^2 E_{\mathrm{THz}}(t) \pm 4\mathrm{i} \omega_{\mathrm{IR}} \dot{E}_{\mathrm{THz}}(t) - \ddot{E}_{\mathrm{THz}}(t) \bigr].
\end{align}
The second origin is the frequency dependence of the third-order susceptibility $\bar{\chi}^{(3)}(\pm \omega_{\mathrm{IR}}, \pm \omega_{\mathrm{IR}}, \omega \mp 2 \omega_{\mathrm{IR}})$.
When $\chi^{(3)}$ is expanded around $\omega = \pm 2\omega_{\mathrm{IR}}$, as in Eqs.~\eqref{eq:chi3_shorthand_p} and \eqref{eq:chi3_shorthand_m}, the linear term in $\omega \mp 2 \omega_{\mathrm{IR}}$ produces a contribution proportional to $\dot{E}_{\mathrm{THz}}(t)$ via inverse Fourier transformation:
\begin{align}
&\int_{-\infty}^{\infty} \frac{\mathrm{d}\omega}{2\pi} \mathrm{e}^{-\mathrm{i}\omega t} \omega^2 (\omega \mp 2\omega_{\mathrm{IR}}) E_{\mathrm{THz}}(\omega \mp 2\omega_{\mathrm{IR}}) \notag \\
&= \mathrm{i} \frac{\mathrm{d}^2}{\mathrm{d}t^2} \int_{-\infty}^{\infty} \frac{\mathrm{d}\omega}{2\pi} \mathrm{e}^{-\mathrm{i}\omega t} \int_{-\infty}^{\infty} \mathrm{d}t' \frac{\mathrm{d} \mathrm{e}^{\mathrm{i}(\omega \mp 2\omega_{\mathrm{IR}})t'}}{\mathrm{d}t'} E_{\mathrm{THz}}(t') \notag \\
&= -\mathrm{i} \frac{\mathrm{d}^2}{\mathrm{d}t^2} \int_{-\infty}^{\infty} \frac{\mathrm{d}\omega}{2\pi} \mathrm{e}^{-\mathrm{i}\omega t} \int_{-\infty}^{\infty} \mathrm{d}t' \mathrm{e}^{\mathrm{i}(\omega \mp 2\omega_{\mathrm{IR}})t'} \dot{E}_{\mathrm{THz}}(t') \notag \\
&= \mathrm{i} \mathrm{e}^{\mp 2\mathrm{i}\omega_{\mathrm{IR}}t} \bigl[ 4 \omega_{\mathrm{IR}}^2 \dot{E}_{\mathrm{THz}}(t) \pm 4\mathrm{i} \omega_{\mathrm{IR}} \ddot{E}_{\mathrm{THz}}(t) - \dddot{E}_{\mathrm{THz}}(t) \bigr],
\end{align}
which is reflected in the terms involving $\bar{\chi}_3'$ in Eq.~\eqref{eq:farfield}.

Since Eq.~\eqref{eq:farfield} takes the form of $E_{\infty}(t) = A(t) \cos(2\omega_{\mathrm{IR}} t) + B(t) \sin(2\omega_{\mathrm{IR}} t)$, where $A(t)$ and $B(t)$ are slowly varying functions of time, the SH intensity is given by the time-averaged square of the field: $\overline{E_{\infty}(t)^2} \approx [A(t)^2 + B(t)^2]/2$.
This leads to the following expressions for the SH intensity in equilibrium,
\begin{align}
&I_{\mathrm{SH}}
= \frac{1}{2} \omega_{\mathrm{IR}}^4 \vert \bar{\chi}_2 \vert^2 E_{\mathrm{IR}}^4, \label{eq:I_SH}
\end{align}
and for the temporal modulation of the SH intensity linearly induced by the pulsed field,
\begin{align}
&\Delta I_{\mathrm{SH}}(t)
= \omega_{\mathrm{IR}}^4 \vert \bar{\chi}_2 \vert E_{\mathrm{IR}}^4 \biggl\{
\vert \bar{\chi}_3 \vert \cos(\phi_2 - \phi_3) E_{\mathrm{THz}}(t) \notag \\
&+ \biggl[ \frac{\vert \bar{\chi}_3 \vert}{\omega_{\mathrm{IR}}} \sin(\phi_2 - \phi_3)
+ \vert \bar{\chi}_3' \vert \sin(\phi_2 - \phi_3') \biggr] \dot{E}_{\mathrm{THz}}(t)
\biggr\}. \label{eq:dI_SH}
\end{align}
Therefore, the relative change in the SH intensity, $\Delta I_{\mathrm{SH}}/I_{\mathrm{SH}}$, can be expressed in the form
\begin{align}
\frac{\Delta I_{\mathrm{SH}}}{I_{\mathrm{SH}}}
= C_0 E_{\mathrm{THz}}(t) + C_1 \frac{\dot{E}_{\mathrm{THz}}(t)}{\omega_{\mathrm{THz}}},
\label{eq:dI_I}
\end{align}
where the coefficients $C_0$ and $C_1 \equiv C_{10} + C_{11}$ are given by
\begin{align}
C_0 &= \frac{2 \vert \bar{\chi}_3 \vert}{\vert \bar{\chi}_2 \vert} \cos(\phi_2 - \phi_3), \label{eq:C0} \\
C_{10} &= \frac{2 \omega_{\mathrm{THz}} \vert \bar{\chi}_3 \vert}{\omega_{\mathrm{IR}} \vert \bar{\chi}_2 \vert} \sin(\phi_2 - \phi_3), \label{eq:C10} \\
C_{11} &= \frac{2 \omega_{\mathrm{THz}} \vert \bar{\chi}_3' \vert}{\vert \bar{\chi}_2 \vert} \sin(\phi_2 - \phi_3'). \label{eq:C11}
\end{align}

A notable result is that, unlike the polarization change $\Delta P(t)$ in Eq.~\eqref{eq:Pt_modulation}, the SH intensity change $\Delta I_{\mathrm{SH}}(t)$ depends not only on $E_{\mathrm{THz}}(t)$ but also on its time derivative $\dot{E}_{\mathrm{THz}}(t)$.
Furthermore, even if the time-derivative component $C_1$ is negligible, the sign of $C_0$ can be negative; that is, $\Delta I_{\mathrm{SH}}(t)$ can be in antiphase with $E_{\mathrm{THz}}(t)$.
Consequently, the commonly assumed relation $\Delta P(t) \propto \Delta I_{\mathrm{SH}}(t)$ does not generally hold.

\section{Numerical analysis} \label{sec:MF}
In this section, we consider a concrete model for ferroelectrics to quantitatively evaluate the contribution of $\dot{E}_{\mathrm{THz}}(t)$ to $\Delta I_{\mathrm{SH}}$.
We introduce the model in Sec.~\ref{sec:model}, followed by an evaluation of its second- and third-order susceptibilities using the spectral representations in Sec.~\ref{sec:susceptibility}.
The real-time simulation method is detailed in Sec.~\ref{sec:simulation}.
We then show the numerical results in Sec.~\ref{sec:results} to verify the theoretical prediction provided by Eq.~\eqref{eq:dI_SH}.

\subsection{Model} \label{sec:model}

\begin{figure*}[t]\centering
\includegraphics[scale=1]{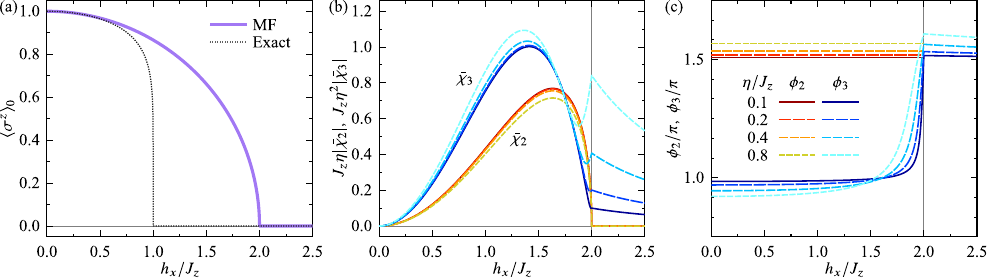}
\caption{(a)~Electric polarization $\langle \sigma^z \rangle_0$ in the ground state.
The dotted curve indicates the exact result, $[1 - (h_x/J_z)^2]^{1/8}$.
(b)~Absolute values and (c) phases of $\bar{\chi}_2$ and $\bar{\chi}_3$ for $\eta/J_z = 0.1$ (solid) to $0.8$ (dashed) and $\omega_{\mathrm{IR}} = E_{\mathrm{g}}/2$.
The phase $\phi_2$ is undefined for $h_x/J_z \geq 2$.
}
\label{fig:gs}
\end{figure*}

We consider the one-dimensional Ising model with a transverse field, whose Hamiltonian is given by
\begin{align}
\mathcal{H}_0 = -J_z \sum_{i} \sigma_i^z \sigma_{i+1}^z -h_x \sum_{i} \sigma_i^x, \label{eq:hamiltonian_tfim}
\end{align}
where $\sigma_i^\alpha\ (\alpha=x,y,z)$ denotes the Pauli matrix at site $i$.
This spin model captures the essential features of electronic ferroelectricity \cite{Ishihara2010, Ikeda2005}: the Ising interaction $J_z$ stabilizes the ferroelectric phase with $\langle \sigma_i^z \rangle_0 \neq 0$, while the transverse field $h_x$ favors the quantum paraelectric phase with $\langle \sigma_i^z \rangle_0 = 0$ \cite{Naka2010, Hotta2010}.
Hereafter, we employ the mean-field (MF) approximation to make the problem tractable, which also enables us to derive explicit expressions for the nonlinear susceptibilities.
The MF Hamiltonian is given by
\begin{align}
\mathcal{H}_{0}^{\mathrm{MF}} = -h_z \sigma^z - h_x \sigma^x, \quad
h_z = 2J_z s_z, \label{eq:hamiltonian_mf}
\end{align}
where the order parameter $s_z = \langle \sigma^z \rangle_0$ is the electric polarization, and the site index has been omitted.
The MF Hamiltonian is diagonalized by a unitary transformation $U_y(\theta)^\dagger \mathcal{H}_{0}^{\mathrm{MF}} U_y(\theta)$ with $U_y(\theta) = \exp(-\mathrm{i}\theta\sigma^y/2)$ and $\theta=\arctan(h_x/h_z)$, yielding
\begin{align}
U_y(\theta)^\dagger \mathcal{H}_{0}^{\mathrm{MF}} U_y(\theta) = \begin{bmatrix} \varepsilon_- & 0 \\ 0 & \varepsilon_+ \end{bmatrix} {}, \quad
\varepsilon_\pm = \pm \sqrt{h_z^2 + h_x^2}.
\end{align}
The ground state and excited state are given by $\vert 0\rangle = U_y(\theta)\vert{\uparrow}\rangle$ and $\vert 1 \rangle = U_y(\theta)\vert{\downarrow}\rangle$, respectively, where $\vert{\uparrow}\rangle$ and $\vert{\downarrow}\rangle$ are the eigenstates of $\sigma^z$ satisfying $\sigma^z \vert{\uparrow}\rangle = \vert{\uparrow}\rangle$ and $\sigma^z \vert{\downarrow}\rangle = - \vert{\downarrow}\rangle$.

The order parameter $s_z$ is determined by the self-consistent equation
\begin{align}
s_z = \langle \sigma^z \rangle_0 = \frac{2J_z s_z}{\sqrt{4J_z^2 s_z^2 + h_x^2}} \label{eq:scf}
\end{align}
at zero temperature.
The solution of Eq.~\eqref{eq:scf} is
\begin{align}
s_z = \begin{cases}
\hfil \sqrt{4J_z^2 - h_x^2} \Big/ (2J_z) & (2J_z>h_x>0) \\
\hfil 0 & (h_x>2J_z>0),
\end{cases}
\label{eq:sz0}
\end{align}
which is plotted in Fig.~\ref{fig:gs}(a), along with the exact solution \cite{Pfeuty1970} for reference.
The ground state is ferroelectric for $h_x/J_z < 2$ and paraelectric for $h_x/J_z > 2$.
The energy gap between $\vert 0 \rangle$ and $\vert 1 \rangle$ can be expressed as
\begin{align}
E_{\mathrm{g}} = \varepsilon_+ - \varepsilon_-
= \begin{cases}
\hfil 4J_z & (2J_z>h_x>0) \\
\hfil 2h_x & (h_x>2J_z>0).
\end{cases}
\end{align}
The matrix elements of the polarization operator $P = \sigma^z$ are
\begin{gather}
\langle 0 \vert P \vert 0 \rangle
= - \langle 1 \vert P \vert 1 \rangle
= \langle \sigma^z \rangle_0,
\label{eq:trans_dipole_00} \\
\langle 0 \vert P \vert 1 \rangle
= \begin{cases}
\hfil -h_x/(2J_z) & (2J_z>h_x>0) \\
\hfil -1 & (h_x>2J_z>0).
\end{cases}
\label{eq:trans_dipole_01}
\end{gather}

\subsection{Nonlinear susceptibility} \label{sec:susceptibility}
Here, we present the explicit forms of the second- and third-order susceptibilities in the ferroelectric ground state, using their spectral representation presented in Appendix.
Assuming that $2\omega_{\mathrm{IR}}$ is resonant with the energy gap $E_{\mathrm{g}} = 4J_z$, and substituting Eqs.~\eqref{eq:sz0}--\eqref{eq:trans_dipole_01} into the spectral representation, we obtain the expression for $\bar{\chi}_2$ in Eq.~\eqref{eq:sh_chi2} as
\begin{align}
\bar{\chi}_2
&= \frac{32 h_x^2 (3J_z + \mathrm{i}\eta) \sqrt{4J_z^2 - h_x^2}}{\mathrm{i}\eta J_z (2J_z - \mathrm{i}\eta) (4J_z + \mathrm{i}\eta) (6J_z + \mathrm{i}\eta) (8J_z + \mathrm{i}\eta)},
\label{eq:chi2_shg}
\end{align}
where $\eta$ is a broadening factor.
This is plotted in Figs.~\ref{fig:gs}(b) and \ref{fig:gs}(c) for $\eta/J_z = 0.1$, $0.2$, $0.4$, and $0.8$ (solid to dashed lines).
As $\eta \to 0$, Eq.~\eqref{eq:chi2_shg} asymptotically reduces to
\begin{align}
\bar{\chi}_2
&\to h_x^2 \sqrt{4J_z^2 - h_x^2} \biggl( \frac{7}{96 J_z^5} -\mathrm{i} \frac{1}{4J_z^4 \eta} \biggr) + \mathcal{O}(\eta^1),
\label{eq:chi2_eta0}
\end{align}
and thus $\vert \bar{\chi}_2 \vert$ scales as $\eta^{-1}$, as shown in Fig.~\ref{fig:gs}(b).
Note that, near the transition point $h_x / J_z = 2$, the polarization $\langle \sigma^z \rangle_0$ is proportional to $\vert \bar{\chi}_2 \vert$, whereas this relation fails for $h_x / J_z \lesssim 1.8$.
This is because the magnitude of the transition dipole moment in Eq.~\eqref{eq:trans_dipole_01} decreases as $h_x$ decreases.

For the third-order susceptibility, since the full forms of $\bar{\chi}_3$ and $\bar{\chi}_3'$ in Eqs.~\eqref{eq:sh_chi3} and \eqref{eq:sh_chi3d} are cumbersome, we present their asymptotic expressions:
\begin{align}
\bar{\chi}_3
&\to -\frac{4 J_z^2 h_x^2 - h_x^4}{4J_z^5 \eta^2} + \mathrm{i} \frac{8J_z^2 h_x^2 - 3 h_x^4}{16 J_z^6 \eta} + \mathcal{O}(\eta^0),
\label{eq:chi3_shg} \\
\bar{\chi}_3'
&\to -\frac{2J_z^2 h_x^2 - h_x^4}{8 J_z^6 \eta^2}
- \mathrm{i} \frac{4J_z^2 h_x^2 - h_x^4}{4J_z^5 \eta^3}
+ \mathcal{O}(\eta^{-1}).
\label{eq:chi3d_shg}
\end{align}
The exact absolute value and phase of $\bar{\chi}_3$ for $\eta/J_z = 0.1$--$0.8$ are also shown in Figs.~\ref{fig:gs}(b) and \ref{fig:gs}(c), which are well approximated by the asymptotic expressions (not shown).
In the ferroelectric phase, $\vert \bar{\chi}_3 \vert$ scales as $\eta^{-2}$ [see Eq.~\eqref{eq:chi3_shg}], while in the paraelectric phase it scales as $\eta^{-1}$.
Additionally, the frequency dependence of $\bar{\chi}^{(3)}(\omega_{\mathrm{IR}}, \omega_{\mathrm{IR}}, \nu)$ in Eq.~\eqref{eq:Ew_inf} is plotted in Fig.~\ref{fig:chi3}, showing a resonance peak at $\nu = 0$.
From Eqs.~\eqref{eq:chi2_eta0}--\eqref{eq:chi3d_shg}, the coefficients in Eqs.~\eqref{eq:C0}--\eqref{eq:C11} can be evaluated as follows:
\begin{align}
C_0
&\to \frac{25h_x^2 - 76J_z^2}{12 J_z^2 \sqrt{4J_z^2 - h_x^2}} + \mathcal{O}(\eta^2), \\
C_{10}
&\to \frac{2\sqrt{4J_z^2 - h_x^2}}{J_z} \frac{\omega_{\mathrm{THz}}}{\omega_{\mathrm{IR}} \eta} + \mathcal{O}(\eta^1), \\
C_{11}
&\to \frac{52 J_z^2 - 19h_x^2}{12 J_z^2 \sqrt{4J_z^2 - h_x^2}} \frac{\omega_{\mathrm{THz}}}{\eta} + \mathcal{O}(\eta^1).
\end{align}
The coefficient $C_0$ vanishes and changes sign at $h_x / J_z = \sqrt{76/25} \approx 1.74$, around which the time-derivative component proportional to $C_1$ is expected to dominate.
This prediction will be numerically verified in Sec.~\ref{sec:results}.

\subsection{Real-time simulation} \label{sec:simulation}
To validate the theoretical predictions presented in Sec.~\ref{sec:theory}, we perform real-time simulations of the MF model in Eq.~\eqref{eq:hamiltonian_mf} when a terahertz pulse is applied.
Here, we describe the methodological details.

\begin{figure}[t]\centering
\includegraphics[scale=1]{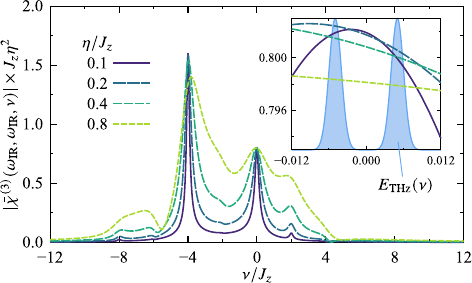}
\caption{Third-order susceptibility $\bar{\chi}^{(3)}(\omega_{\mathrm{IR}}, \omega_{\mathrm{IR}}, \nu = \omega-2\omega_{\mathrm{IR}})$ for $\eta/J_z = 0.1$ to $0.8$, with $h_x/J_z = 1.7$ and $\omega_{\mathrm{IR}} = E_{\mathrm{g}}/2$.
The inset provides an enlarged view, where the shaded blue curve represents $E_{\mathrm{THz}}(\nu)$.
}
\label{fig:chi3}
\end{figure}

The density matrix $\rho(t)$ evolves according to the von Neumann equation with a relaxation-time approximation:
\begin{align}
\dot{\rho}(t) = -\mathrm{i}[\mathcal{H}(t),\rho(t)] - \varGamma [\rho(t)-\rho_0],
\label{eq:vonNeumann}
\end{align}
where $\varGamma$ is the relaxation rate, and $\rho_0 = \vert 0 \rangle\langle 0 \vert$ represents the density matrix of the ground state.
The time-dependent Hamiltonian is given by
\begin{align}
\mathcal{H}(t) &= -h_z(t) \sigma^z - h_x \sigma^x, \\
h_z(t) &= 2J_z \langle \sigma^z \rangle_0 + E(t),
\end{align}
where $E(t)$ is the total electric field from Eq.~\eqref{eq:Et}, comprising a continuous wave (IR probe) and an off-resonant terahertz pump pulse:
\begin{align}
E(t) &= E_{\mathrm{IR}} \cos(\omega_{\mathrm{IR}}t) + E_{\mathrm{THz}}(t), \\
E_{\mathrm{THz}}(t) &= E_{\mathrm{THz}}^0 \exp\biggl(-\frac{t^2}{2t_{\mathrm{w}}^2}\biggr) \cos(\omega_{\mathrm{THz}}t).
\end{align}
Hereafter, we set $E_{\mathrm{IR}} = 0.005J_z$, $\omega_{\mathrm{IR}} = E_{\mathrm{g}}/2 = 2J_z$, $E_{\mathrm{THz}}^0 = 10^{-5}J_z$, $\omega_{\mathrm{THz}} = 0.005J_z$, and $t_{\mathrm{w}} = 1000J_z^{-1}$.
The Fourier spectrum of the terahertz pulse is shown in the inset of Fig.~\ref{fig:chi3}, along with $\bar{\chi}^{(3)}(\omega_{\mathrm{IR}}, \omega_{\mathrm{IR}}, \nu)$ for various $\eta$.
For the parameters adopted, we observe that $\bar{\chi}^{(3)}(\omega_{\mathrm{IR}}, \omega_{\mathrm{IR}}, \nu)$ remains nearly constant over the spectral range of $E_{\mathrm{THz}}(\nu)$, which justifies the linear approximation employed in Eqs.~\eqref{eq:chi3_shorthand_p} and \eqref{eq:chi3_shorthand_m}.
In the simulations presented in Sec.~\ref{sec:results}, we focus on the case with $\varGamma = \eta = 0.1 J_z$, which corresponds to the worst-case scenario for the linear approximation accuracy among the $\eta$ values shown in Fig.~\ref{fig:chi3}.

\begin{figure*}[t]\centering
\includegraphics[scale=1]{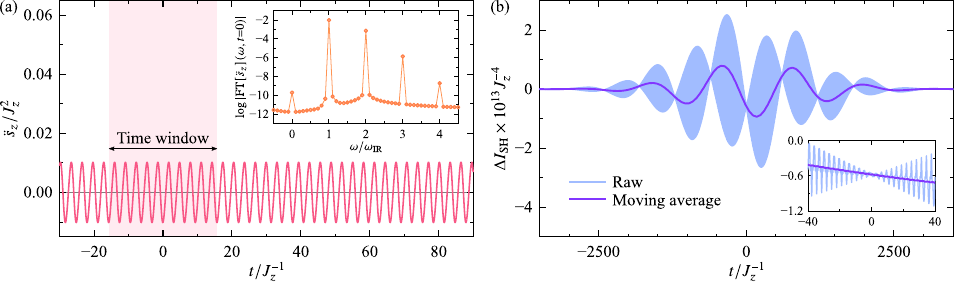}
\caption{(a)~Second derivative of the polarization, $\ddot{s}_z(t)$.
The inset shows its Fourier amplitude within the shaded time window centered at $t = 0$.
(b)~Change in the SH intensity, $\Delta I_{\mathrm{SH}}(t)$.
The inset is an enlarged view around $t = 0$.
The blue and the bold purple curves indicate the raw data and the moving-average data, respectively.
The transverse field strength is set to $h_x / J_z = 1.74$ in (a) and (b).
}
\label{fig:time_ft}
\end{figure*}

To evaluate the SH response, we first compute the second derivative of the polarization,
\begin{align}
\ddot{s}_{z}(t) = \Tr[\ddot{\rho}(t) \sigma^z],
\end{align}
where $\ddot{\rho}(t)$ is explicitly obtained by differentiating Eq.~\eqref{eq:vonNeumann}:
\begin{gather}
\ddot{\rho}(t)
= -\mathrm{i}[\dot{\mathcal{H}}(t),\rho(t)]
-\mathrm{i}[\mathcal{H}(t),\dot{\rho}(t)]
- \varGamma \dot{\rho}(t), \\
\dot{\mathcal{H}}(t) = - \dot{E}(t) \sigma^z.
\end{gather}
Figure~\ref{fig:time_ft}(a) shows an example of the temporal profile of $\ddot{s}_z(t)$.
Since $E_{\mathrm{IR}} \gg E_{\mathrm{THz}}^0$, a sinusoidal oscillation at the probe frequency $\omega_{\mathrm{IR}}$ is predominantly observed.
We then extract the SH amplitude by applying a windowed Fourier transformation to $\ddot{s}_z(t)$.
Specifically, we slide a time window of width $T = 2\pi M/\omega_{\mathrm{IR}}$, where $M$ is an integer, and then compute the discrete Fourier transformation:
\begin{align}
\mathrm{FT}[\ddot{s}](\omega,t)
&= \frac{2\Delta t}{T} \mathrm{e}^{\mathrm{i} \omega (t-T/2)} \sum_{n=0}^{\frac{T}{\Delta t} - 1} \mathrm{e}^{\mathrm{i} \omega n\Delta t} \ddot{s}_z(t - T/2 + n\Delta t),
\label{eq:ft}
\end{align}
where $\Delta t$ is the fixed time step used in the fourth-order Runge--Kutta integration of Eq.~\eqref{eq:vonNeumann}.
To ensure that the frequency grid contains exactly the integer multiples of the probe frequency $\omega_{\mathrm{IR}}$, we set
\begin{align}
\Delta t = \frac{T}{2N_{\mathrm{max}} M},
\end{align}
with $N_{\mathrm{max}}$ denoting the highest positive harmonic order to be resolved.
The prefactor $2\Delta t/T$ in Eq.~\eqref{eq:ft} is chosen such that $\vert \mathrm{FT}[\ddot{s}](\omega, t) \vert = A$ when $\ddot{s}_z(t) = A \cos(\omega t)$.
In our calculations, we choose $M = 10$ and the smallest integer $N_{\mathrm{max}} = 158$ such that $\Delta t \leq 0.01 J_{z}^{-1}$.
This approach based on the discrete Fourier transformation was also adopted in a previous study of high harmonic generation \cite{Ono2024}.

From the Fourier spectrum in Eq.~\eqref{eq:ft}, we extract the SH amplitude at $\omega = 2\omega_{\mathrm{IR}}$ and compute the time-averaged SH intensity as
\begin{align}
I_{\mathrm{SH}} + \Delta I_{\mathrm{SH}}(t)
= \frac{\vert \mathrm{FT}[\ddot{s}](2\omega_{\mathrm{IR}}, t) \vert^2}{2}.
\end{align}
Since this quantity exhibits rapid oscillations [see the blue curve in Fig.~\ref{fig:time_ft}(b)], we apply a moving average over a time span of $4\pi/\omega_{\mathrm{IR}}$, thereby obtaining a slowly varying SH intensity [the bold purple curve in Fig.~\ref{fig:time_ft}(b)] that directly corresponds to Eq.~\eqref{eq:dI_SH}.

\subsection{Numerical results} \label{sec:results}

\begin{figure*}[t]\centering
\includegraphics[scale=1]{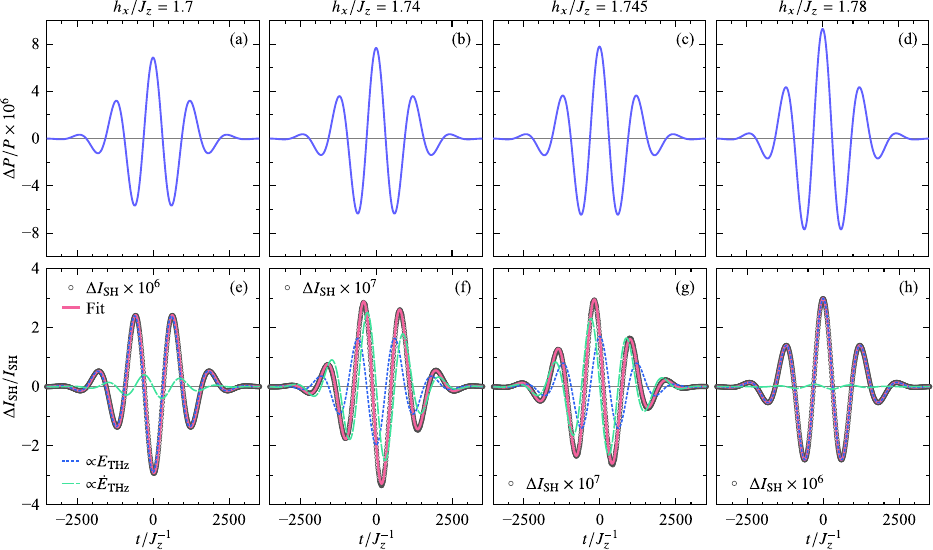}
\caption{Relative changes in (a)--(d) the electric polarization $\Delta P/P$ and (e)--(h) the SH intensity $\Delta I_{\mathrm{SH}}/I_{\mathrm{SH}}$, for $h_x/J_z = 1.7$, $1.74$, $1.745$, and $1.78$ (left to right).
The solid red curves represent fitted results, along with their $E_{\mathrm{THz}}(t)$ component (dotted blue) and $\dot{E}_{\mathrm{THz}}(t)$ component (dashed green).
}
\label{fig:sh}
\end{figure*}

Figure~\ref{fig:sh} shows the temporal changes in polarization $\Delta P(t)$ and SH intensity $\Delta I_{\mathrm{SH}}(t)$.
The temporal profiles of $\Delta P/P$ in Figs.~\ref{fig:sh}(a)--\ref{fig:sh}(d) closely follow the waveform of the terahertz pulse, $E_{\mathrm{THz}}(t)$, and their amplitude increases with increasing $h_x$.
By contrast, the changes in the SH intensity are more pronounced.
For $h_x/J_z = 1.78$ [Fig.~\ref{fig:sh}(h)], $\Delta I_{\mathrm{SH}}(t)$ is nearly in phase with $E_{\mathrm{THz}}(t)$, whereas for $h_x/J_z = 1.7$ [Fig.~\ref{fig:sh}(e)], it is in antiphase.
At intermediate values $h_x/J_z = 1.74$ and $1.745$ [Figs.~\ref{fig:sh}(f) and \ref{fig:sh}(g)], $\Delta I_{\mathrm{SH}}(t)$ exhibits a distinct phase shift relative to $E_{\mathrm{THz}}(t)$.
The red lines in Figs.~\ref{fig:sh}(e)--\ref{fig:sh}(h) represent the fits to Eq.~\eqref{eq:dI_I}, where $C_0$ and $C_1$ are treated as fitting parameters.
These results indicate that, in addition to the component proportional to $E_{\mathrm{THz}}(t)$ (dotted blue lines), a significant contribution proportional to $\dot{E}_{\mathrm{THz}}(t)$ (dashed green lines) emerges despite $\omega_{\mathrm{THz}} \ll \omega_{\mathrm{IR}}$, as predicted in Sec.~\ref{sec:theory}.

\begin{figure*}[t]\centering
\includegraphics[scale=1]{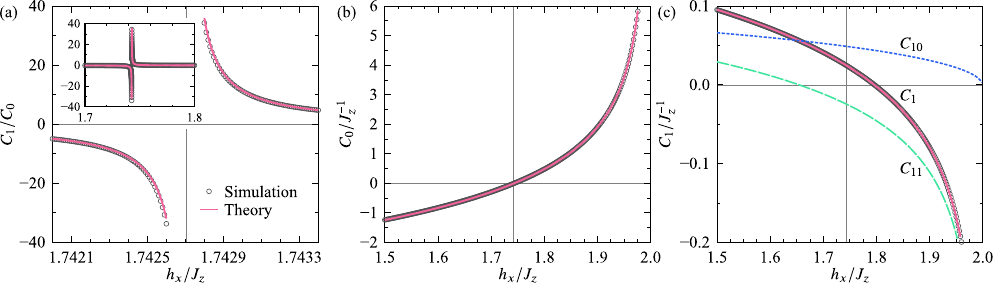}
\caption{(a)~Ratio $C_1/C_0$ representing the relative contribution of the time-derivative component.
The inset shows a broader view for $h_x/J_z \in [1.7, 1.8]$.
The circular markers show the results of the real-time simulations, and the solid red curve indicates the theoretical values calculated from Eqs.~\eqref{eq:C0}--\eqref{eq:C11}.
[(b) and (c)]~Weight of the component proportional to (b) $E_{\mathrm{THz}}$ and (c) $\dot{E}_{\mathrm{THz}}/\omega_{\mathrm{THz}}$.
The gray vertical lines in (a)--(c) indicate $h_x/J_z = 1.742\,706\,2$, at which $C_0$ vanishes.
}
\label{fig:ratio}
\end{figure*}

We performed the same fitting procedure for various values of $h_x$ and plotted the ratio of the time-derivative component, $C_1 / C_0$, as a function of $h_x$ in Fig.~\ref{fig:ratio}(a).
The black circles are simulation results obtained with a relaxation rate of $\varGamma/J_z = 0.1$, while the solid red line represents the analytical prediction from Eqs.~\eqref{eq:C0}--\eqref{eq:C11} using a broadening factor $\eta/J_z = 0.1$.
The two results agree remarkably well, validating our theoretical framework.
As expected from the asymptotic analysis in Sec.~\ref{sec:susceptibility}, $C_1 / C_0$ diverges near $h_x/J_z \approx 1.74$.
This divergence occurs because $C_0$ vanishes at $h_x/J_z \approx 1.7427$, as shown in Fig.~\ref{fig:ratio}(b).
Figure~\ref{fig:ratio}(c) plots $C_1$ along with its components $C_{10}$ and $C_{11}$.
In the region where $C_1/C_0$ diverges, $C_1$ remains nonzero, and contributions from $C_{10}$ and $C_{11}$ are comparable in magnitude.
The divergences of $C_0$ and $C_{11}$ as $h_x / J_z \to 2$ arise because $\vert\bar{\chi}_2\vert$ approaches zero.

\section{Discussion} \label{sec:discussion}
The most significant assumption in our theory is the linear approximation of the third-order susceptibility, as given in Eqs.~\eqref{eq:chi3_shorthand_p} and \eqref{eq:chi3_shorthand_m}.
Provided this approximation holds, Eq.~\eqref{eq:dI_SH} offers a quantitative estimate of the SH intensity change and should be valid even in more complex systems, including real materials.
In the MF transverse-field Ising model discussed in Sec.~\ref{sec:MF}, by considering the $h_x$ dependence of Eqs.~\eqref{eq:trans_dipole_00} and \eqref{eq:trans_dipole_01}, we suppose that $C_0$ vanishes when the magnitude of the polarization expectation value $\langle 0 \vert P \vert 0 \rangle$ becomes comparable to that of the transition dipole moment $\langle 0 \vert P \vert 1 \rangle$.
However, the universality of this behavior is still an open question.
It is therefore important to apply our formulas to other systems to determine whether the time-derivative component can dominate and whether $\Delta I_{\mathrm{SH}}(t)$ remains in phase with $E_{\mathrm{THz}}(t)$.

If one extends beyond the linear approximation to include second-order terms, Eq.~\eqref{eq:Ew_inf} yields contributions of the form $\omega^2 (\omega \mp 2\omega_{\mathrm{IR}})^2 E_{\mathrm{THz}}(\omega \mp 2\omega_{\mathrm{IR}})$, which in the time domain correspond to second- and higher-order time derivatives of $E_{\mathrm{THz}}(t)$.
Experimental detection of these contributions would be highly challenging with current experimental capabilities.
Additionally, if $\Delta I_{\mathrm{SH}}$ depends nonlinearly on $E_{\mathrm{THz}}$, the present formulation breaks down, and a higher-order perturbative expansion becomes necessary.
However, there is no reason to expect that the simple relation $\Delta P \propto \Delta I_{\mathrm{SH}}$ holds in such strongly nonlinear regimes.

In this work, we have discussed the modulation of the polarization and the SH intensity by an external electric field pulse.
Other intriguing second-order phenomena, such as optical rectification and the bulk photovoltaic effect, are also described by the second-order response function.
Similarly, the linear modulation of these effects can be described by the third-order susceptibility.
Although in Eqs.~\eqref{eq:Pw2} and \eqref{eq:Pw3} we focused on contributions around $\omega \approx \pm 2\omega_{\mathrm{IR}}$, an analogous discussion could address the response near $\omega \approx 0$.
However, when considering a monochromatic probe as in Eq.~\eqref{eq:Ew}, the far field $E_{\infty}$ in Eq.~\eqref{eq:dipole_w} vanishes at $\omega = 0$, and thus the theory would require some modifications.
A detailed investigation of this aspect is left for future work.

\section{Summary} \label{sec:summary}
We investigated the temporal modulation of the electric polarization $\Delta P(t)$ and second harmonic intensity $\Delta I_{\mathrm{SH}}(t)$, both of which are linearly induced by a low-frequency electric field pulse $E_{\mathrm{THz}}(t)$.
Using nonlinear response theory, we derived general expressions for $\Delta P$ and $\Delta I_{\mathrm{SH}}$, demonstrating that $\Delta I_{\mathrm{SH}}$ can contain terms proportional not only to $E_{\mathrm{THz}}(t)$ but also to its time derivative $\dot{E}_{\mathrm{THz}}(t)$.
Through an analysis of the transverse-field Ising model within the mean-field approximation, we showed that for most parameter regimes, $\Delta I_{\mathrm{SH}}$ is predominantly proportional to $E_{\mathrm{THz}}(t)$, in agreement with common assumptions.
We also found that the time-derivative component can become significant under certain conditions, and our simulations quantitatively agree with the theoretical predictions.
Furthermore, we showed that when the time-derivative component is negligible, $\Delta I_{\mathrm{SH}}$ can nonetheless be in antiphase with $E_{\mathrm{THz}}(t)$; that is, $\Delta I_{\mathrm{SH}}$ can decrease while $\Delta P$ increases, and vice versa.
These findings reveal that the commonly assumed relation $\Delta P(t) \propto \Delta I_{\mathrm{SH}}(t)$ does not always hold, thereby enabling a more accurate analysis of the ultrafast dynamics in ferroelectrics driven by electric field pulses.

\begin{acknowledgments}
The author is grateful to Shinichiro Iwai, Hirotake Itoh, Yoichi Okimoto, and Shin-ya Koshihara for their insightful comments in the early stages of this work.
This work was supported by JSPS KAKENHI Grants No.\ JP23K13052, No.\ JP23K25805, and No.\ JP24K00563.
The numerical calculations were performed using the facilities of the Supercomputer Center, the Institute for Solid State Physics, the University of Tokyo.
\end{acknowledgments}

\appendix*
\renewcommand{\theequation}{A\arabic{equation}}
\section{Spectral representation of nonlinear susceptibility} \label{sec:spectralrep}
We present the spectral representation of the second- and third-order susceptibilities \cite{Liu2022}.
Let $\mathcal{H}_0$ be the unperturbed Hamiltonian, with its many-body eigenstates $\{ \vert n \rangle \}$ and eigenenergies $\{ \varepsilon_n \}$.
The ground state of $\mathcal{H}_0$ is denoted by $\vert 0 \rangle$.
For generality, we consider a set of physical quantities $\{ Q^{\alpha} \}$ that are linearly coupled to external fields $\{ f_{\beta}(t) \}$ as
\begin{align}
\mathcal{V}(t) = - \sum_{\beta} Q^\beta f_{\beta}(t).
\end{align}
The nonlinear response function is generally expressed as
\begin{align}
&\chi_{\alpha\beta_1 \cdots \beta_n}^{(n)}(t;t_1,t_2,\dots,t_n) \notag \\
&= \mathrm{i}^n \varTheta(t-t_1)\varTheta(t_1-t_2)\cdots\varTheta(t_{n-1}-t_n) \notag \\
&\quad\times \left\langle [[\cdots [[\hat{Q}^{\alpha}(t),\hat{Q}^{\beta_1}(t_1)],\hat{Q}^{\beta_2}(t_2)],\cdots ], \hat{Q}^{\beta_n}(t_n)] \right\rangle_0.
\label{eq:chi_general}
\end{align}
By inserting the resolution of the identity, $1 = \sum_n \vert n \rangle \langle n \vert$, into Eq.~\eqref{eq:chi_general} and performing the Fourier transformation, we obtain the expression for the second-order response function in the ground state as
\begin{align}
&\chi_{\alpha\beta_1\beta_2}^{(2)}(\omega_1, \omega_2) \notag \\*
&= \sum_{mn} \Biggl[
\frac{Q^{\alpha}_{0m} Q^{\beta_1}_{mn} Q^{\beta_2}_{n0}}{(\omega_1^+ + \omega_2^+ - \varepsilon_m + \varepsilon_0) (\omega_2^+ - \varepsilon_n + \varepsilon_0)} \notag \\
&\quad +
\frac{Q^{\beta_2}_{0m} Q^{\beta_1}_{mn} Q^{\alpha}_{n0}}{(\omega_1^+ + \omega_2^+ - \varepsilon_0 + \varepsilon_n) (\omega_2^+ - \varepsilon_0 + \varepsilon_m)} \notag \\
&\quad -
\frac{Q^{\beta_2}_{0m} Q^{\alpha}_{mn} Q^{\beta_1}_{n0}}{(\omega_1^+ + \omega_2^+ - \varepsilon_n + \varepsilon_m) (\omega_2^+ - \varepsilon_0 + \varepsilon_m)} \notag \\
&\quad -
\frac{Q^{\beta_1}_{0m} Q^{\alpha}_{mn} Q^{\beta_2}_{n0}}{(\omega_1^+ + \omega_2^+ - \varepsilon_n + \varepsilon_m) (\omega_2^+ - \varepsilon_n + \varepsilon_0)}
\Biggr],
\label{eq:chi2_spectral}
\end{align}
with $Q_{mn}^{\alpha} = \langle m \vert Q^{\alpha} \vert n \rangle$.
Similarly, the spectral representation of the third-order response function is given by
\begin{align}
&\chi_{\alpha\beta_1\beta_2\beta_3}^{(3)}(\omega_1, \omega_2, \omega_3) \notag \\*
&= \sum_{lmn} \biggl[
- \frac{Q^{\alpha}_{0l} Q^{\beta_1}_{lm} Q^{\beta_2}_{mn} Q^{\beta_3}_{n0}}{(\omega_1^+ + \omega_2^+ + \omega_3^+ - \varepsilon_{l0}) (\omega_2^+ + \omega_3^+ - \varepsilon_{m0}) (\omega_3^+ - \varepsilon_{n0})} \notag \\
&\quad +
\frac{Q^{\beta_3}_{0l} Q^{\beta_2}_{lm} Q^{\beta_1}_{mn} Q^{\alpha}_{n0}}{(\omega_1^+ + \omega_2^+ + \omega_3^+ - \varepsilon_{0n}) (\omega_2^+ + \omega_3^+ - \varepsilon_{0m}) (\omega_3^+ - \varepsilon_{0l})} \notag \\
&\quad +
\frac{Q^{\beta_1}_{0l} Q^{\alpha}_{lm} Q^{\beta_2}_{mn} Q^{\beta_3}_{n0}}{(\omega_1^+ + \omega_2^+ + \omega_3^+ - \varepsilon_{ml}) (\omega_2^+ + \omega_3^+ - \varepsilon_{m0}) (\omega_3^+ - \varepsilon_{n0})} \notag \\
&\quad +
\frac{Q^{\beta_2}_{0l} Q^{\alpha}_{lm} Q^{\beta_1}_{mn} Q^{\beta_3}_{n0}}{(\omega_1^+ + \omega_2^+ + \omega_3^+ - \varepsilon_{ml}) (\omega_2^+ + \omega_3^+ - \varepsilon_{nl}) (\omega_3^+ - \varepsilon_{n0})} \notag \\
&\quad +
\frac{Q^{\beta_3}_{0l} Q^{\alpha}_{lm} Q^{\beta_1}_{mn} Q^{\beta_2}_{n0}}{(\omega_1^+ + \omega_2^+ + \omega_3^+ - \varepsilon_{ml}) (\omega_2^+ + \omega_3^+ - \varepsilon_{nl}) (\omega_3^+ - \varepsilon_{0l})} \notag \\
&\quad -
\frac{Q^{\beta_3}_{0l} Q^{\beta_2}_{lm} Q^{\alpha}_{mn} Q^{\beta_1}_{n0}}{(\omega_1^+ + \omega_2^+ + \omega_3^+ - \varepsilon_{nm}) (\omega_2^+ + \omega_3^+ - \varepsilon_{0m}) (\omega_3^+ - \varepsilon_{0l})} \notag \\
&\quad -
\frac{Q^{\beta_3}_{0l} Q^{\beta_1}_{lm} Q^{\alpha}_{mn} Q^{\beta_2}_{n0}}{(\omega_1^+ + \omega_2^+ + \omega_3^+ - \varepsilon_{nm}) (\omega_2^+ + \omega_3^+ - \varepsilon_{nl}) (\omega_3^+ - \varepsilon_{0l})} \notag \\
&\quad -
\frac{Q^{\beta_2}_{0l} Q^{\beta_1}_{lm} Q^{\alpha}_{mn} Q^{\beta_3}_{n0}}{(\omega_1^+ + \omega_2^+ + \omega_3^+ - \varepsilon_{nm}) (\omega_2^+ + \omega_3^+ - \varepsilon_{nl}) (\omega_3^+ - \varepsilon_{n0})}
\biggr],
\label{eq:chi3_spectral}
\end{align}
where $\varepsilon_{mn} = \varepsilon_{m} - \varepsilon_{n}$.
In Eqs.~\eqref{eq:chi2_spectral} and \eqref{eq:chi3_spectral}, $\omega_i^+$ denotes a complex frequency with a positive imaginary part.
It was shown in Ref.\ \cite{Ono2025} that applying the following substitution yields good agreement between the nonlinear responses obtained from real-time simulations and those derived from the spectral representation:
\begin{align}
\omega_3^+ &\to \omega_3 + \mathrm{i} \eta, \\
\omega_2^+ + \omega_3^+ &\to \omega_2 + \omega_3 + \mathrm{i} \eta, \\
\omega_1^+ + \omega_2^+ + \omega_3^+ &\to \omega_1 + \omega_2 + \omega_3 + \mathrm{i} \eta,
\end{align}
where $\eta$ is set to the relaxation rate in Eq.~\eqref{eq:vonNeumann}, i.e., $\eta = \varGamma$.
Additionally, the nonlinear susceptibilities in Eqs.~\eqref{eq:chi2_spectral} and \eqref{eq:chi3_spectral} need to be symmetrized in accordance with Eq.~\eqref{eq:chi_symmetrized}.

\bibliography{reference}

\end{document}